\documentclass[aps,prl,reprint,twocolumn,superscriptaddress,showpacs,showkeys,a4paper]{revtex4-1}

\usepackage{amsmath,amssymb,amstext}
\usepackage[squaren]{SIunits}
\usepackage[usenames,dvipsnames]{color}
\usepackage{graphicx,subfigure,relsize}
\usepackage{bm,bbold,braket}

\definecolor{mygreen}{rgb}{0,0.5,0}
\definecolor{myblue}{rgb}{0,0,0.75}
\definecolor{mymagenta}{cmyk}{0,1,0,0.12}

\renewcommand{\braket}[1]{\langle #1 \rangle}
\newcommand{\ave}[1]{\langle #1 \rangle}
\newcommand{\ri}{r_{i}}
\newcommand{\Jai}{J_{\alpha,i}}
\newcommand{\Jaj}{J_{\alpha,j}}
\newcommand{\jai}{j_{\alpha,i}}
\newcommand{\jain}{\jai^{(n)}}
\newcommand{\Jx}{J_x}
\newcommand{\Jy}{J_y}
\newcommand{\Jz}{J_z}
\newcommand{\Jzi}{J_{z,i}}
\newcommand{\Ja}{J_{\alpha}}
\newcommand{\Jb}{J_{\beta}}
\newcommand{\Jc}{J_{\gamma}}
\newcommand{\Sa}{S_{\alpha}}
\newcommand{\Sx}{S_1}
\newcommand{\Sy}{S_2}
\newcommand{\Sz}{S_3}
\newcommand{\sa}{\sigma_{\alpha}}
\newcommand{\aap}{a_{+}}
\newcommand{\aam}{a_{-}}
\newcommand{\aapm}{a_{\pm}}
\newcommand{\Na}{n_{\rm a}}
\newcommand{\Ns}{n_{\rm s}}
\newcommand{\Nph}{N_{\rm ph}}
\newcommand{\NA}{N_{\rm A}}
\newcommand{\Op}{\Omega_{p}}
\newcommand{\Hp}{H_{p}}
\newcommand{\kp}{k_{p}}
\newcommand{\kapp}{C_{p}}
\newcommand{\Kapp}{C_{p}^{2}}
\newcommand{\tG}{\tilde{\Gamma}}
\newcommand{\GG}{\Gamma}
\newcommand{\Gaa}{\tG_{\alpha\alpha}}
\newcommand{\Gab}{\tG_{\alpha\beta}}

\newcommand{\tGaa}{\GG_{\alpha\alpha}}
\newcommand{\Rm}{R_{\rm m}}
\newcommand{\Rn}{R_{\rm n}}
\newcommand{\Gmn}{\tG_{\rm mn}}

\newcommand{\Gzz}{\tG_{zz}}
\newcommand{\tGzz}{\GG_{zz}}
\newcommand{\supin}{^{\rm (in)}}
\newcommand{\supout}{^{\rm (out)}}
\newcommand{\supM}{^{\rm (M)}}
\newcommand{\PP}{\Pi_{\rm 2}}
\newcommand{\MP}{{\rm MP}}
\newcommand{\fm}{f_{p}}
\newcommand{\gm}{g_{p}}
\newcommand{\etam}{\eta_{p}}
\newcommand{\Cdes}{C_{\rm des}(\delta r)}
\newcommand{\eabc}{\epsilon_{\alpha\beta\gamma}}

\begin{document}

\title{Quantum control of spin-correlations in ultracold lattice gases}

\author{P.~Hauke}
    \email{philipp.hauke@icfo.es}
    \affiliation{ICFO-Institut de Ciencies Fotoniques, Av. Carl Friedrich Gauss, 3, 08860 Castelldefels, Barcelona, Spain.}
\author{R.J.~Sewell}
    \affiliation{ICFO-Institut de Ciencies Fotoniques, Av. Carl Friedrich Gauss, 3, 08860 Castelldefels, Barcelona, Spain.}
\author{M.W.~Mitchell}
    \affiliation{ICFO-Institut de Ciencies Fotoniques, Av. Carl Friedrich Gauss, 3, 08860 Castelldefels, Barcelona, Spain.}
    \affiliation{ICREA-Instituci\'{o} Catalana de Recerca i Estudis Avan\c{c}ats, 08015 Barcelona, Spain}
\author{M.~Lewenstein}
    \affiliation{ICFO-Institut de Ciencies Fotoniques, Av. Carl Friedrich Gauss, 3, 08860 Castelldefels, Barcelona, Spain.}
    \affiliation{ICREA-Instituci\'{o} Catalana de Recerca i Estudis Avan\c{c}ats, 08015 Barcelona, Spain}

\date{\today}

\begin{abstract}
We demonstrate that it is possible to prepare a lattice gas of ultracold atoms with a desired non-classical spin-correlation function using atom-light interaction of the kind routinely employed in quantum spin polarization spectroscopy. Our method is based on quantum non-demolition (QND) measurement and feedback, and allows in particular to create on demand exponentially or algebraically decaying correlations, as well as a certain degree of multi-partite entanglement. 
\end{abstract}



\maketitle

Ultra-cold atomic gases trapped in optical lattices offer an unprecedented playground for studying the quantum phases of many-body systems~\cite{Bloch2008,*Lewenstein2007,*Esslinger2010,*Lewenstein2012}. In particular, quantum states of ultra-cold lattice gases with spin degrees of freedom may be used to simulate quantum magnetism and to investigate physics relevant for our understanding of high-$T_c$ superconductivity~\cite{Anderson1987,*Kotliar1988,*Lee2006}. While enormous progress has been made towards engineering such systems, achieving the regime of high-$T_c$ superconductivity remains experimentally extremely challenging because of the low temperatures required~\cite{Lewenstein2012}.

\begin{figure}[t]
    \centering
    \includegraphics[width=\columnwidth]{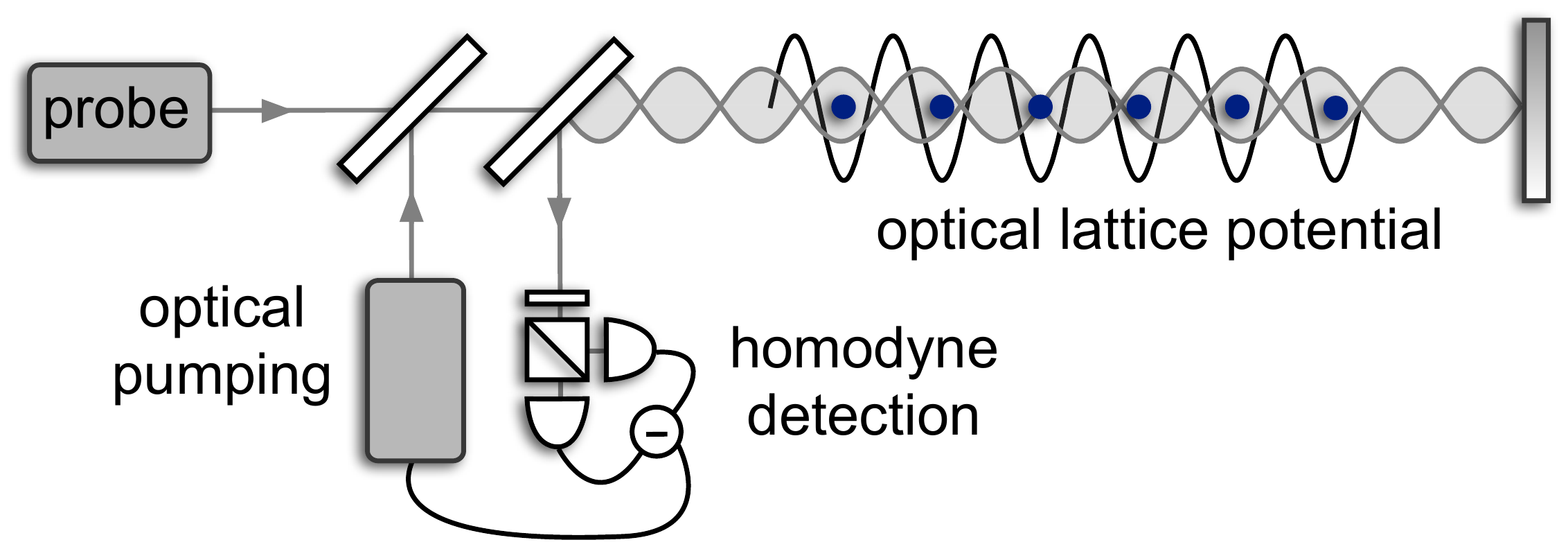}%
    \caption{{\bf Proposed experimental set-up.}  Atoms trapped in an optical lattice (black) are probed with a far-detuned, linearly polarized standing-wave light pulse with wavevector $\kp$.  Afterwards, the probe beam is outcoupled to a homodyne detector, where $\Sy$ is recorded.  A projection-noise limited measurement induces spin squeezing, introducing quantum correlations among the atoms in spatial mode $k=2\kp$.  Feedback is applied via optical pumping to set $\ave{\Ja(k)}=0$.  Successive spin components $\Ja$ can then be separately squeezed by coherently rotating the atomic spin between measurements.  The procedure is repeated for a set of wavevectors $\kp$, with interaction strengths $\Op$ weighted by the corresponding amplitude of the cosine Fourier transform of the desired spatial correlation signature.\label{fig:expt}}
\end{figure}

In this context, quantum spin polarization spectroscopy (SPS)~\cite{Eckert2007} has emerged as a promising technique for \emph{detecting} quantum phases in lattice gases via the coherent mapping of spin-correlations onto scattered light in a quantum non-demolition (QND) measurement.  In particular, spatially-resolved SPS that employs standing-wave laser configurations~\cite{Eckert2008} allows direct probing of magnetic structure factors and order parameters~\cite{Roscilde2009,*De-Chiara2011,*Weitenberg2011,*De-Chiara2011a,*Meineke2012}.  In this manuscript, we propose inverting the SPS scheme in order to \emph{prepare} a lattice gas with a desired non-classical spin correlation function. Motivated by the experimental demonstration of spin-squeezing via QND measurements~\cite{Appel2009,*Takano2009,*Schleier-Smith2010,*Chen2011,*Sewell2011}, and by the recent extension of these ideas to unpolarized ensembles~\cite{Toth2010}, we demonstrate that a simple modification of the experimental scheme of Ref.~\cite{Eckert2008} (illustrated in Fig.~\ref{fig:expt}) allows for the on-demand preparation of lattice gases with arbitrary spin-correlation functions. 


\paragraph{Atom-light interaction.---}
We consider the interaction of atoms trapped in a one-dimensional optical lattice potential with a set of standing-wave pulses of near-resonant light with wave-numbers $\kp$.  The atoms are described by collective variables $\Jai\equiv\sum_{n=1}^{\Na}\jain$, where the index $n$ runs over the $\Na$ atoms at lattice site $i$ and $\alpha=x,y,z$ labels the components of the atomic spin operators with length $j$.  With $\Ns$ lattice sites, the total number of atoms is $\NA=\Ns\Na$. The photons are described by collective Stokes operators $\Sa$ with $\alpha=1,2,3$, defined as $\Sa\equiv\dfrac{1}{2}(\aap^{\dagger},\aam^{\dagger})\sa(\aap,\aam)^T$, where the $\sa$ are the Pauli matrices, and $\aapm$ are annihilation operators for the spatial and temporal mode of the pulse with circular plus/minus polarization.  The atom--light interaction for a single pulse is then described by the effective Hamiltonian~\cite{Hammerer2004,*Echaniz2008} 
\begin{equation} \Hp=\Op\sum_{i=1}^{\Ns}c_{i}(\kp)\Jzi\Sz\,,
    \label{eq:ham}
\end{equation}
where $c_{i}(\kp)=(1+\cos(2\kp\ri))/2$ describes the standing-wave intensity profile, and the coupling constants $\Op$ depend on the probe detuning and intensity.  Eq.~\eqref{eq:ham} describes a QND measurement that induces spin-squeezing of the $\Jz$ component of the collective atomic mode $\Ja(k)\equiv\sum_{i=1}^{\Ns}\Jai\exp(ik\ri)/\sqrt{\Ns}$ with $k=\pm2\kp$.  For multi-level alkali atoms, this effective Hamiltonian can be synthesized using multicolor or dynamical-decoupling probing techniques~\cite{Appel2009,Koschorreck2010b}.

\paragraph{Measurement \& feedback---}
We model the interaction using methods developed for treating the Gaussian dynamics of collective-variable systems~\cite{Molmer2004,Madsen2004,Koschorreck2009,Toth2010}, with the assumption that $\Na\gg1$. The full system is then described by the operators $\Rm(k)=\{\Jx(k),\Jy(k),\Jz(k),\Sx,\Sy,\Sz\}$ and the covariances $\Gmn(k_1,k_2)\equiv\ave{\Rm(k_1)\Rn(k_2)+\Rn(k_2)\Rm(k_1)}/2-\ave{\Rm(k_1)}\ave{\Rn(k_2)}$. The dynamical equations for the covariances can be derived from the Heisenberg equation of motion for the operators, where, in the small-angle regime, an operator changes as $\Rm(k)\supout=\Rm(k)\supin-i\tau[\Rm(k)\supin,\Hp]$, where $\tau$ is the pulse duration.  We assume that the atomic and light variables are initially uncorrelated, and that the atomic covariances $\Gab\supin(k_1,k_2)=0$ $\forall \alpha\neq\beta$.  For simplicity we also assume a uniform atomic filling factor.  

For an input $\Sx$-polarized pulse, the only covariances that change due to the pulse are 
\begin{eqnarray}
\label{eq:covout}
&&  \tG_{z2}\supout(k) =\frac{\kapp}{2\sqrt{j}} \Gzz\supin(\kp),\text{ and }   \\
&&  \tG_{22}\supout =\tG_{22}\supin +\frac{\Kapp}{8j}\left[ 2\Gzz\supin(0) + \Gzz\supin(2\kp) \right.\\ \notag
&&  \qquad\qquad    + \left. \Gzz\supin(-2\kp)\right] \,,
\end{eqnarray}
where the coupling strength $\kapp=\tau\Op\Sx\sqrt{\Ns j/S_{0}}$ and we define $\Gzz(k)\equiv \Gzz(0,k)+\Gzz(2\kp,k)/2+\Gzz(-2\kp,k)/2$.

Detection of $\Sy$ then transfers the correlations described in Eqs.~\eqref{eq:covout} to the atoms.  This can be modeled as a projection $\Gamma\supM=\Gamma\supout-\Gamma\supout(\PP\Gamma\supout\PP)^\MP\Gamma\supout$, where $\MP$ indicates the Moore-Penrose pseudoinverse and $\PP=\text{diag}(0,0,0,0,1,0)$~\cite{Toth2010}.  After the measurement, the atomic covariances are
\begin{equation}
    \Gab\supM(k_1,k_2)=\Gab\supout(k_1,k_2)-\frac{\Gamma_{\alpha2}\supout(k_1)\Gamma_{2\beta}\supout(k_2)}{\Gamma_{22}\supout}.
\label{eq:atomcov}
\end{equation}
Eqns.~\eqref{eq:covout} and~\eqref{eq:atomcov} imply that if the covariance matrix $\Gab$ is initially diagonal, then the only atomic covariances changed by the interaction--measurement process are $\Gzz(k_1,k_2)$, i.e. the measurement induces spin squeezing of the $\Jz(2\kp)$ mode. Furthermore, the process is highly symmetric, preserving $\Gamma_{\alpha\alpha}(k,k')=\Gamma_{\alpha\alpha}(k,-k')$ for $k\neq k'$ and $\Gamma_{\alpha\alpha}(k,k')=\Gamma_{\alpha\alpha}(k',k)$ $\forall k,k'$.

The orthogonal spin components $\Ja(2\kp)$ can be successively measured by coherently rotating the atomic spin between measurements.  To allow the measurement-induced squeezing to be repeated for each spin component, we require $\ave{\Ja(2\kp)}=0$, which allows us to avoid measurement induced back-action due to the Heisenberg uncertainty relation $(\Delta\Ja(k_1))^2(\Delta\Jb(k_2))^2\ge\eabc\tfrac{1}{4\Ns}{|\ave{\Jc(k_1+k_2)}|^2}$.
To obtain $\ave{\Ja(2\kp)}=0$, the result of the measurement of $\Sy$ can be used as the input to an optical pumping feedback process: a weak pulse of near-resonant light at wavevector $\kp$ with an intensity proportional to $\Sy\supout$ will set $\braket{\Jz(2\kp)+\Jz(-2\kp)}=0$, and a second pulse with a half-period phase shift sets $\braket{\Jz(2\kp)-\Jz(-2\kp)}=0$, so that $\braket{\Jz(2\kp)}=\braket{\Jz(-2\kp)}=0$.  As shown in Ref.~\cite{Toth2010}, this feedback introduces spin noise $\propto N^{1/4}$, which is negligible in the thermodynamic limit~\footnote{Alternatively, data with $\left<\Ja(2\kp)\right>\ll\GG^{(0)}$ $\forall$ could be post-selected based on the measurement outcomes}.



\begin{figure*}[t]
    \centering
    \includegraphics[width=\textwidth]{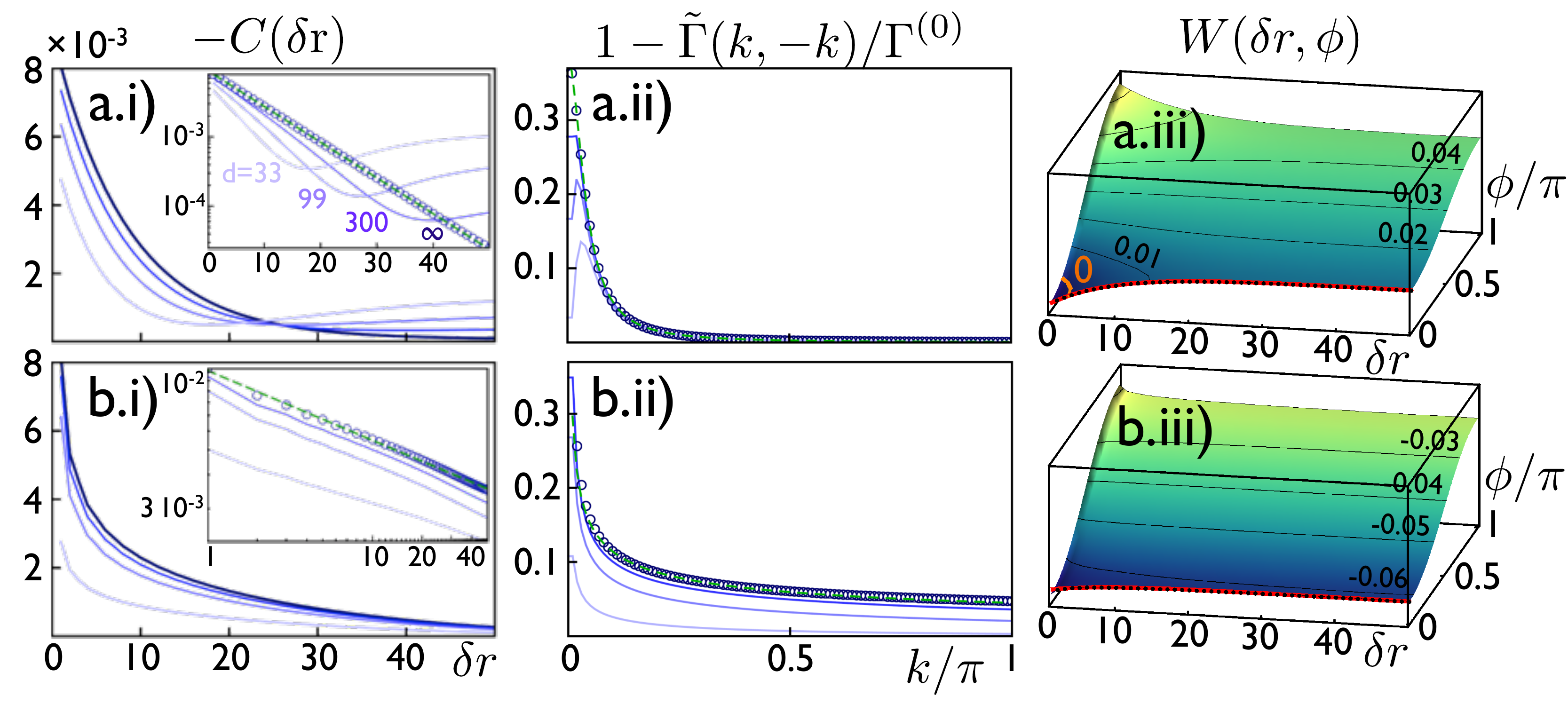}
    \caption{\textbf{Numerical results for (a) exponential decay, and (b) algebraic decay.}  \textbf{(i) First column:}  Real-space spin correlation functions.  The lines from darker to brighter shades of blue (and in the main panels from top to bottom) are for optical depth $d=\infty$, 300, 99 and 33.  The $d=\infty$ data is plotted with open circles to allow comparison with the fitted curves (green dashed lines).  The insets in {\bf (a.i)} and in {\bf (b.i)} are log-linear and log-log plots, respectively, where for clarity we subtract $C(\delta r\to\infty)$, extracted from a fit.  In {\bf (a.i)}, without decoherence, the decay follows an exponential fit.  Deviations at large distance become stronger with increasing decoherence (decreasing $d$), and $C(\delta r\to\infty)$ cannot be reliably determined, yielding deviations from straight lines.  In {\bf (b.i)}, the curves are straight lines for all values of $d$, and algebraic fits are very accurate.  \textbf{(ii) Second column:} The covariances $\Gamma(k,-k)$ closely follow the cosine Fourier-transform of the desired correlation signature $\Cdes$, even for small $d$.  Deviations occur only primarily at small $k$ where the pulse strength $\kapp$ is high.  \textbf{(iii) Third column:}  If the entanglement witness $W(\delta r,\phi)$ is negative, the system diplays non-classical correlations.  In {\bf (a.iii)} and {\bf (b.iii)}, the red lines are fits to $W(\delta r,0)$ with exponential and algebraic decay respectively.  Note that the orange contour line in {\bf (a.iii)} marks $W(\delta r,\phi)=0$.
\label{fig:expalgGaussAndexpalgGaussEntanglement}}
\end{figure*}

\paragraph{Strategy---}
We now motivate a strategy that exploits the spatial dependence induced by $c_{i}(\kp)$ in Eq.~\eqref{eq:ham} to systematically manipulate the real-space spin--spin correlation function
\begin{eqnarray}
\label{eq:spinspin}
    \tGaa(r_1,r_2)&\equiv&\ave{\Ja(r_1)\Ja(r_2)}-\ave{\Ja(r_1)}\ave{\Ja(r_2)}   \\
    &=&\frac{1}{\Ns}\sum_{k_1,k_2=1}^{\Ns}\exp(ik_1r_1)\exp(ik_2r_2)\Gaa(k_1,k_2)\,    \notag\nonumber
\end{eqnarray}
where we label the $k$-vectors in the first Brillouin zone (BZ) from 1 to $\Ns$.  Eqs.~(\ref{eq:covout}) and~(\ref{eq:atomcov}) imply that the covariances of a given collective mode $\Ja(k)$ are only altered by a pulse with $k=2\kp$~\footnote{The pulse with $k=2\kp$ also introduces small correlations to the zero mode $\Gamma_{zz}(k,0)\neq0$, which results in a change of $\tGzz(k,k)$ and $\tGzz(k,-k)$ by subsequent pulses with $k'\neq k$.  Below we demonstrate that this has negligible effect on the outcome.}.  This suggests that we can manipulate $\tGaa(r_1,r_2)$ with a sequence of pulses with wavevectors $2\kp$ that cover the first BZ.  Now we show that this can be done by an appropriate choice of coupling constants $\kapp$.

We assume that we start with a completely mixed initial state with covariances $\tGzz^{(0)}(r_1,r_2)=\GG^{(0)}\delta_{r_1,r_2}$ where $\GG^{(0)}=\Na j(j+1)/3$.  Under the approximation that the covariances $\tGaa(k_1,k_2)$ with $k_2\neq\pm k_1,0$ remain small, Eq.~\eqref{eq:spinspin} becomes 
\begin{widetext}
\begin{align}
\label{eq:GzzCosFourierTransform}
    \tGzz(r_1,r_2)&\approx\frac{1}{\Ns}\Gzz(0,0)+\frac{2}{\Ns}\sum_{p}\left[\cos(2\kp r_1)+\cos(2\kp r_2)\right]\Gzz(2\kp,0)\nonumber\\
                &+\frac{2}{\Ns}\sum_{p}\left[\cos(2\kp(r_1+r_2))\Gzz(2\kp,2\kp) + \cos(2\kp(r_1-r_2))\Gzz(2\kp,-2\kp)\right].
\end{align}
\end{widetext}
The spatial dependence is strongly dominated by the $\Gzz(2\kp,-2\kp)$ term, which, after the pulse with $\kp$, changes as $\Gzz\supout(2\kp,-2\kp)=\GG^{(0)}(1-\fm/4)$, where we define the scaled coupling constants 
\begin{equation}
    \fm\equiv\frac{\GG^{(0)}\Kapp}{4j\Gamma_{22}\supout}.
\label{eq:fm}
\end{equation}

Assuming that these covariances are not changed by subsequent pulses~\footnote{This restriction could be relaxed with a more sophisticated strategy that takes into account the change in the correlations at one wavevector due to squeezing a different wavevector.}, Eq.~\eqref{eq:GzzCosFourierTransform} becomes
\begin{align}
        \tGzz(r_1,r_2)&\approx\frac{\Gzz(0,0)}{\Ns} \\
    &+\frac{2}{\Ns}\sum_{p}\cos(2\kp(r_1-r_2))\GG^{(0)}\left(1-\frac{\fm}{4}\right). \notag
\end{align}
The spatial dependence is given by the final term, which is the cosine Fourier transform of the $\fm$.  This suggests the following strategy for manipulating the spin--spin correlations $\tGzz(r_1,r_2)$: Let $\tGzz(r_1,r_1+\delta r)=\Cdes$ be the desired output correlation signature.  To determine the coupling strength $\kapp$ that should be used for each wavevector $\kp$ in order to create $\Cdes$, we approximate $\fm$ by the inverse cosine Fourier transform of $-4\Cdes/\GG^{(0)}$. Further, in Eq.~\eqref{eq:fm}, we replace the covariances $\Gzz(k_1,k_2)$ in the expression for $\Gamma_{22}\supout$ with the completely mixed values $\GG^{(0)}$. Both approximations are valid for realistic experimental parameters.
Now, we can solve Eq.~\eqref{eq:fm} for $\kapp$,
\begin{equation}
    \kapp=2\sqrt{j}\sqrt{\frac{\Gamma_{22}\supin\fm}{1-\gm\GG^{(0)}\fm}}
\label{eq:k}
\end{equation}
where $\gm=\tfrac{9}{2}$ for $\kp=0$ and $\gm=\tfrac{3}{2}$ otherwise.

The coupling strengths $\kapp$ can be adjusted experimentally by choosing detuning $\Delta$, intensity, and duration of the pulse appropriately. In fact, $\kapp=\sqrt{\NA\Nph}\sigma\gamma/A\Delta$, where $\sigma$ is the on-resonance cross section for the probe transition, $\gamma$ the spontaneous decay rate, and $A$ the cross section of the atomic ensemble illuminated by the probe. With a finite on-resonance optical depth $d$, $\kapp$ is related to the probability of spontaneous emission $\etam$ via $\kapp=\sqrt{d\etam}$, giving a trade-off between coupling strength and decoherence. Decoherence due to spontaneous excitation by the probe pulse is included in the model, following Refs.~\cite{Giedke2002,Toth2010,*Madsen2004,*Koschorreck2009}, by updating the atomic covariances according to $\Gaa^{(\eta)}(k_1,k_2)=(1-2\etam)\Gaa(k_1,k_2)+2\etam\Gamma^{(0)}\delta_{k_1,-k_2}$.

\paragraph{Entanglement witness.---}
A special kind of correlation, entanglement, is particularly important in the context of quantum information processing and many-body systems~\cite{Osborne2002,*Osterloh2002,*Jozsa2003,*Verstraete2004,*Guhne2009}. To show that our proposal can create multipartite entanglement, we derive an entanglement witness for the multimode spatial correlations induced by the procedure described above. 
Generalizing the strategy of Refs.~\cite{Krammer2009,*Cramer2011,*De-Chiara2011b}, we use the witness $W\equiv\mathcal{S}/\Na-1$, such that $W<0$ implies entanglement, where we define 
\begin{equation}
    \mathcal{S}\equiv\sum_{\alpha}\mathcal{S}_{\alpha}=\sum_{\alpha}\sum_{i,j=1}^{\Ns}\ave{\Jai\Jaj}f^{*}(r_i)f(r_j)\,.
\label{eq:s}
\end{equation}
Here, $f(r_i)$ is any normalized function $\sum_{i=1}^{\Ns}\left|f(r_i)\right|^2=1$.  This definition encompasses and generalizes the plane waves described in Refs.~\cite{Krammer2009,Cramer2011}, and allows us to calculate the entanglement witness $W$ as a function of spatial separation, which may be of general interest outside this particular example.

To probe spatial dependence, we calculate the entanglement between two sets of lattice bins $r_{s=1\ldots m}$ and $r_{w=1\ldots n}$ separated by a distance $\delta r$ using the witness $W$ with the function
\begin{equation}
    f(r_i)=
    \begin{cases}
        1 & \text{if $r_i\in r_s$,} \\
        \exp(i\phi) & \text{if $r_i\in r_w$,}   \\
        0 & \text{otherwise.}
    \end{cases}
\end{equation}
For given $\delta r$, $W$ can be minimized with respect to $\phi$.

\paragraph{Numerical results.---}
We illustrate this technique using a 1D chain of spin $j=1$ atoms with $\Ns=200$ sites and $\Na=10$ atoms per site~\footnote{Note that the same results generalize to a single atom per site, as long as we bin the atoms into $\Ns$ bins with $\Na$ atoms per bin, and redefine the coupling constant $\kapp$ as an average over the $\Na$ atoms in each bin.}, which is related to the bilinear-biquadratic Hamiltonian, which has a rich phase diagram displaying ferromagnetic, critical, dimerized, and Haldane phases, each with distinctive spatial correlation signatures~\cite{Haldane1983,Fath1991,Bursill1995,Schollwock1996,Imambekov2003,Buchta2005,Rizzi2005,Lauchli2006}.  We have also checked that our method can be used to prepare correlation signatures of more exotic quantum phases, such as the  critical phase of the bilinear-biquadratic Hamiltonian, which has a structure factor peaked at $k=\pm2\pi/3$~\cite{Bursill1995,*Schollwock1996}, and algebraically decaying correlations with characteristic period-3 oscillations~\cite{Fath1991}.

We demonstrate the preparation of spin correlations $\Cdes$ with: (a) an exponential decay $\exp(-r/\xi)$ with a correlation length $\xi$, corresponding to gapped phases, such as spin liquids which are conjectured to appear in the vicinity of high-$T_c$ superconductivity~\cite{Anderson1987,Kotliar1988,Lee2006,Misguich2004,*Lhuillier2005}; and (b) an algebraic decay $r^{-\zeta}$, corresponding to critical phases and quantum critical points of the phase diagram.  We illustrate case (a) with $\xi=5$ and case (b) with $\zeta=0.7$. We compute the $\fm$ corresponding to $\Cdes$ as described above, apply the pulses in sequence to the atoms, and compute the resulting real-space spin-correlations 
\begin{equation}
    C(\delta r)=\frac{1}{\Ns/4}\sum_{i=1}^{\Ns/4}\sum_{\alpha}\tGaa(r_i,r_i+\delta r)/\GG^{(0)}
\end{equation}
after all pulses have been applied. The only remaining free parameter is then the maximum coupling strength ${\rm max}_p\{\kapp\}$, which we set to $\approx0.95$, ensuring that the approximations suggesting the used coupling strengths~\eqref{eq:k} are valid~\footnote{This is a conservative choice: we could increase $\kapp$ and calculate the $\fm$ taking into account the change in the covariances $\Gzz(k_1,k_2)$ after each pulse in the sequence.}.


The numerical results are shown in Fig.~\ref{fig:expalgGaussAndexpalgGaussEntanglement}.  At large optical depth, the desired correlation signatures $\Cdes$ coincide well with the calculated correlation function in both cases (first column).  For case (a), the exponential decay is maintained over several orders of magnitude, and fits to the short-range behaviour yield a correlation length close to the desired $\xi=5$.  For case (b), a clear algebraic decay is seen with a fitted $\zeta\approx0.4$.  Deviations from the desired parameters induced by finite optical depth could be further compensated by adjusting the $\kapp$ appropriately.  The real-space correlation signature can also be extracted by fitting the Fourier transform of $\Cdes$ to the covariances $\tG(k,-k)$ (second column), which are the observables that are measured and manipulated in the experiment.  Finally, we calculate the entanglement witness $W$ for a single bin entangled with a chain of 106 bins (third column).  In both cases, $W$ is minimized for $\phi=0$ and decays exponentially (algebraically) with $\delta r$ following the spatial behavior of the spin correlation function.  Notably, the entanglement is stronger for algebraic decay.

\paragraph{Outlook.---}We have demonstrated that, with a simple modification of the experimental scheme discussed in Ref.~\cite{Eckert2008}, it is possible to engineer a quantum lattice gas with an arbitrary non-classical spin correlation function.  We have illustrated the procedure with two examples mimicking the quantum phases of the bilinear-biquadratic Hamiltonian, demonstrating how to prepare exponentially- and algebraically-decaying correlations. We have also checked that the method can be extended to the spatial correlations signatures of more exotic quantum phases. We have generalized the entanglement witness proposed in Refs.~\cite{Krammer2009,*Cramer2011} and shown that the engineered  spin-correlations entail multimode atomic entanglement.  In our calculations, we make conservative assumptions about the experimental parameters, leaving considerable scope for further optimization of the procedure, which is readily extendible to higher dimensions and larger-spin systems, with both fermionic or bosonic atoms.


\begin{acknowledgments}
This work was supported by the Spanish MINECO (project MAGO, FIS2011-23520), the Spanish MICINN (TOQATA, FIS2008-00784), Catalunya-Caixa, by the ERC (AQUMET, QUAGATUA), and EU projects AQUTE and NAMEQUAM. 
\end{acknowledgments}




%

\end{document}